\documentclass[preprint,12pt]{elsarticle}

\usepackage[T1]{fontenc}
\usepackage{amsmath,amssymb,amsthm}
\usepackage{graphicx}
\usepackage{booktabs}
\usepackage{xcolor}
\usepackage[hidelinks]{hyperref}

\newtheorem{theorem}{Theorem}
\newtheorem{proposition}{Proposition}
\newtheorem{corollary}{Corollary}
\newtheorem{lemma}{Lemma}
\theoremstyle{definition}
\newtheorem{definition}{Definition}
\newtheorem{assumption}{Assumption}
\newtheorem{example}{Example}
\theoremstyle{remark}
\newtheorem{remark}{Remark}

\newcommand{\tv}{\tilde{v}}
\newcommand{\R}{\mathbb{R}}

\journal{Operations Research Letters}

\makeatletter
\def\ps@acceptedfirst{%
  \let\@oddhead\@empty
  \let\@evenhead\@empty
  \def\@oddfoot{\parbox[t]{\textwidth}{\centering\footnotesize
    Accepted manuscript. The published version appears in Operations Research Letters,
    \url{https://doi.org/10.1016/j.orl.2026.107581}\\
    \copyright\ 2026. This manuscript version is made available under the CC-BY-NC-ND 4.0 license
    \url{https://creativecommons.org/licenses/by-nc-nd/4.0/}}}%
  \let\@evenfoot\@oddfoot}
\makeatother

\begin{document}

\begin{frontmatter}

\title{A maximum-entropy reduction for coalition formation\\ with anonymous outsiders\tnoteref{ded}}
\tnotetext[ded]{In memory of Dimitris Kouvelis.}

\author{Paraskevas V. Lekeas\corref{cor1}}
\ead{plekeas@gmail.com}
\cortext[cor1]{Corresponding author. Paraskevas V. Lekeas, Morton Grove, IL, USA.}
\address{Morton Grove, IL, USA}

\begin{abstract}
A coalition contemplating deviation must predict how the $m$ outsiders reorganise, a problem with Bell-number many outcomes. Under anonymity, own-size worths, and equal division, we reduce prediction to choosing among $m$ block sizes: the size whose stationarity hyperplane lies nearest the worth vector, computable in $O(m)$ after $O(m^2)$ preprocessing. The reference is the size-biased occupancy of a uniform partition; the rule admits a replicator-stationarity reading and is benchmarked against the welfare optimum, classical core conjectures, and belief-based Cournot cores.
\end{abstract}


\end{frontmatter}
\thispagestyle{acceptedfirst}

\section{Introduction}\label{sec:intro}

Consider a group of players who are weighing whether to leave a grand coalition. Whether the move pays depends on something they do not control, namely what the players left behind will do. In a cooperative game in partition function form [1], this dependence is explicit, as the worth a coalition can secure varies with the coalition structure [2] formed by the remaining players. A deviating coalition $S$ therefore faces a prediction problem before it faces a payoff problem. The $m$ players outside $S$ can reorganise into any of $B_m$ partitions, where the Bell number $B_m$ [3] grows superexponentially, and selecting a worth-maximising structure is NP-hard for general characteristic functions [4], a difficulty that has generated a substantial algorithmic literature [5, 6].

This note studies the prediction problem under three restrictions, stated formally in Section~\ref{sec:model}: nobody's identity matters (\emph{anonymity}), a block's worth depends only on its own size (\emph{own-size worths}), and blocks divide worth \emph{equally}. Two consequences should be separated at the outset. First, these assumptions significantly reduce the computational obstruction. Payoff-relevant outcomes collapse from the $B_m$ labelled partitions to the integer partitions of $m$. Their number still grows subexponentially, but the welfare-optimal outsider structure is computable by an elementary dynamic program in $O(m^2)$ time. So hardness is not the issue in our setting. What the assumptions do not supply is a point prediction, namely a principled answer to a simple question: around which of the $m$ block sizes will the uncoordinated outsiders settle? This note proposes such an answer.

Our idea can be stated in one breath: assume nothing about the outsiders beyond the rules of the game, compute what an average, unprejudiced arrangement would earn per head, and predict the block size whose condition of earning exactly that average is nearest to being satisfied. Formally, we propose selecting the size whose \emph{stationarity hyperplane} lies nearest the observed worth vector. The construction rests on the maximum-entropy prior over outsider organisations, namely the uniform distribution over the $B_m$ partitions. This prior induces a size occupancy $\pi$: the distribution of the block size of a randomly chosen outsider (Definition~\ref{def:pi}), which Proposition~\ref{prop:ref} characterises in closed form. A size is called stationary when its per-capita worth equals the population average at $\pi$. The stationarity conditions of the $m$ sizes turn out to be $m$ hyperplanes through the origin of worth space (Theorem~\ref{thm:main}). The prediction is then a single nearest-hyperplane computation, costing $O(m)$ per query after $O(m^2)$ preprocessing (Corollary~\ref{cor:classifier}). The rule and a companion ranking by raw residuals have exact operational meanings, developed in Remark~\ref{rem:interp}. A size's residual is the per-capita side payment that would bring it to the agnostic average. Its hyperplane distance is the smallest perturbation of the worth vector that would make it stationary. So the companion selects the size cheapest to stabilise by side payments and the rule the size cheapest to make stationary. We also highlight an evolutionary reading, in which the residuals are instantaneous per-capita growth rates of the replicator dynamics [7, 8] evaluated at $\pi$, so the companion selects the least-disturbed share and the rule the size closest to stationarity in worth space. Throughout, the note supplies a principled selection under the agnostic prior; behavioural foundations that would elevate it to a forecast are left open in Section~\ref{sec:conc}. Section~\ref{sec:bench} benchmarks the rule against the welfare optimum on five worth families, against baselines that embody the classical core conjectures about outsider behaviour (Remark~\ref{rem:cores}), reporting unfavourable as well as favourable comparisons. Section~\ref{sec:cournot} confronts the rule with the belief-based cores of [9] in Cournot oligopolies, where the sufficiency threshold for substitutes grows as $\Theta(\sqrt n)$ while, under the self-consistency construction used there, the prediction tracks the prior's typical block size. Section~\ref{sec:conc} states open problems.

\section{Model and preliminaries}\label{sec:model}

Let $N=\{1,\dots,n\}$ be the player set and let $S\subset N$ with $|S|=s$ be a coalition weighing a deviation. Two roles recur throughout and we fix the words now. The \emph{deviators} are the members of $S$; they make the prediction. The \emph{outsiders} are the $m=n-s\ge 1$ players of $N\setminus S$; they are the ones being predicted. A \emph{partition} $P=\{b_1,\dots,b_r\}$ of $N\setminus S$ collects the outsiders into disjoint nonempty blocks, and together with the deviating block it forms the coalition structure $Q_P=\{S\}\cup P$ of the full player set $N$. The number of outsider partitions is the Bell number $B_m$ [3], satisfying $B_m=\sum_{i=0}^{m-1}\binom{m-1}{i}B_i$ with $B_0=1$. In partition function form [1], each block $b\in Q_P$ obtains a worth that may depend on the entire structure, written $u(b;Q_P)$; our object of study is the outsiders' side of this map, the worths of the blocks $b\in P$. Three assumptions now shrink this generality to the symmetric case we study. In plain terms they say that names do not matter, that only a block's own size matters, and that blockmates share equally.

\begin{assumption}[Anonymity]\label{a:anon}
Worths depend only on sizes: for every outsider block $b\in P$ one has $u(b;Q_P)=v\bigl(|b|;\lambda(Q_P)\bigr)$ for some function $v$, whose first argument is the block's size and whose second argument $\lambda(Q_P)$ is the multiset of block sizes of the structure. No player identity enters.
\end{assumption}

\begin{assumption}[Own-size worths]\label{a:ownsize}
The function $v$ of Assumption~\ref{a:anon} does not depend on its second argument: $v(j;\lambda)=v(j)$ for every size profile $\lambda$. We write $v(j)$ for this common value, the worth of any block of $j$ outsiders, and set $v(0)=0$.
\end{assumption}

\begin{assumption}[Equal division]\label{a:equal}
A member of a size-$j$ block receives the per-capita worth $f_j := v(j)/j$.
\end{assumption}

\begin{remark}[What anonymity does and does not remove]\label{rem:anon}
Assumption~\ref{a:anon} removes the identities of players, not the externalities between blocks: under Assumption~\ref{a:anon} alone, a block's worth still varies with the entire size profile. Assumption~\ref{a:ownsize} is the stronger restriction that removes cross-block externalities among the outsiders. The externality that motivates the prediction problem runs one way and survives both assumptions. The deviating coalition's own worth still depends on how the outsiders organise, while outsider blocks' worths do not depend on one another. Sections~\ref{sec:rule} and~\ref{sec:bench} operate under Assumptions~\ref{a:anon} to~\ref{a:equal}; Section~\ref{sec:cournot} treats an anonymous game in which Assumption~\ref{a:ownsize} fails, namely Cournot oligopoly, and shows how a self-consistency device reduces it to own-size form.
\end{remark}

Under Assumptions~\ref{a:anon} to~\ref{a:equal}, everything the deviators can observe is the \emph{worth vector} $v=(v(1),\dots,v(m))\in\R^m$, a point of \emph{worth space} whose $j$th entry is the worth of any size-$j$ outsider block. The deviators' own size $s$ is fixed, which fixes the number $m=n-s$ of outsiders, and nothing is predicted about $S$. What the deviators must produce is a prediction of how the $m$ outsiders organise, that is, an integer partition of $m$ listing the sizes of the blocks they form. The rule of Section~\ref{sec:rule} does not search over all such partitions. It selects a single block size $\hat k\in\{1,\dots,m\}$, and the following convention completes $\hat k$ into one specific integer partition of $m$.

\begin{definition}[Tiling convention]\label{def:tiling}
Given a block size $\hat k\in\{1,\dots,m\}$, the \emph{predicted structure} tiles the $m$ outsiders into $\lfloor m/\hat k\rfloor$ blocks of size $\hat k$ and, when $\rho=m\bmod\hat k$ is positive, one further block of size $\rho$. It has $\lceil m/\hat k\rceil$ blocks, and its per-capita worth is $\bigl(\lfloor m/\hat k\rfloor\,v(\hat k)+v(\rho)\bigr)/m$, with $v(0)=0$ as stipulated. For instance, $m=27$ and $\hat k=4$ give six blocks of four and one of three.
\end{definition}

Saying that the outsiders organise around size $\hat k$ means that they form the structure of Definition~\ref{def:tiling}. The rule of Section~\ref{sec:rule} will read this task through an evolutionary lens, so we first set up the requisite vocabulary: population states, fitness, and the replicator dynamics.

\paragraph{Population states and the replicator equation}
Rather than tracking who sits with whom, we describe an organisation by a cruder statistic, namely how many outsiders sit in blocks of each size. Identify an outsider organisation with its \emph{occupancy} $x\in\Delta^{m-1}=\{x\in\R^m:x_k\ge0,\ \sum_k x_k=1\}$, where $x_k$ is the fraction of outsiders sitting in size-$k$ blocks. A concrete partition $P$ induces the occupancy $x_k(P)=k\,n_k(P)/m$, where $n_k(P)$ is the number of blocks of $P$ of size $k$. Next comes a notion of payoff. Treat the choice of a block size as a strategy; under Assumptions~\ref{a:ownsize} and~\ref{a:equal} its \emph{fitness} is the per-capita worth $f_k=v(k)/k$, a constant that does not depend on the state $x$. With states and fitnesses in place, the replicator equation [7, 8] is
\begin{equation}\label{eq:replicator}
\dot x_k \;=\; x_k\bigl(f_k-\bar f(x)\bigr), \qquad \bar f(x)=\sum\nolimits_{j=1}^{m}x_j f_j ,
\end{equation}
where $\bar f(x)$ is the \emph{mean fitness} at state $x$. The equation has a simple reading. A size whose members earn more than the current average grows; one whose members earn less shrinks.

\begin{definition}[Stationarity of a size]\label{def:stat}
Size $k$ is \emph{stationary at state $x$} if $f_k=\bar f(x)$, meaning its per-capita worth equals the population average, so that by Equation~\eqref{eq:replicator} its share is momentarily constant.
\end{definition}

Stationarity is relative to a state, and the deviators, by hypothesis, do not know which state the outsiders occupy. The natural resolution is to evaluate it at the state that assumes the least.

\paragraph{The agnostic prior}
What does assuming the least mean exactly? For a probability distribution $q$ over the $B_m$ outsider partitions it means maximum Shannon entropy $\mathcal{H}(q)=-\sum_{\omega} q(\omega)\log q(\omega)$, the sum running over the partitions, and by Jensen's inequality the maximiser is the uniform distribution [10]. The phrase ``uniform distribution'' requires fixing the space it lives on. We place it on the $B_m$ labelled partitions, the finest description of who joins whom; the exchangeability motivated by Assumption~\ref{a:anon} makes arrangements within a size profile equally plausible, and maximising entropy then weights each profile by its number of arrangements, recovering uniformity over the labelled partitions. Probability distributions over the outsiders' partitions were used as coalitional beliefs in Cournot games in [11]; here the belief is derived from the entropy principle and developed into a general prediction rule. Uniformity over size profiles is an alternative agnostic prior. The form of the results below does not depend on the choice, as only the averaging weights, introduced shortly in Proposition~\ref{prop:avg}, would change; individual predictions can differ across the two priors.

One step remains before the vocabulary is complete. The prior lives on partitions, while the dynamics of Equation~\eqref{eq:replicator} live on states, and the object that carries the maximum-entropy prior into state space is the following.

\begin{definition}[Maximum-entropy occupancy]\label{def:pi}
$\pi_k$ is the probability that a uniformly chosen outsider lies in a size-$k$ block when the partition is drawn uniformly from the $B_m$ partitions of $N\setminus S$.
\end{definition}

\section{The nearest-hyperplane rule}\label{sec:rule}

As noted in Remark~\ref{rem:anon}, Assumptions~\ref{a:anon} to~\ref{a:equal} are in force throughout this section and Section~\ref{sec:bench}. We now construct the prediction rule, in three steps: average the worths over the agnostic prior (Proposition~\ref{prop:avg}), recognise the average as the replicator mean fitness at $\pi$ (Proposition~\ref{prop:ref}), and classify a worth vector by its distance to the resulting stationarity conditions (Theorem~\ref{thm:main} and Definition~\ref{def:rule}).

\begin{proposition}[Average per-capita worth]\label{prop:avg}
The average per-capita worth over all $B_m$ partitions of $N\setminus S$, with equal weight, is
\begin{equation}\label{eq:avg}
\begin{split}
\tv&=\frac{1}{m\,B_m}\sum_{j=1}^{m} v(j)\binom{m}{j}B_{m-j}
=\sum_{j=1}^m w_j\,v(j),\\
w_j&=\frac{1}{m\,B_m}\binom{m}{j}B_{m-j}.
\end{split}
\end{equation}
\end{proposition}

\begin{proof}
A partition's per-capita worth is $\frac1m\sum_b v(|b|)$ over its blocks $b$. Summing over all partitions and grouping by block size $j$, the number of (partition, distinguished size-$j$ block) pairs is $\binom{m}{j}B_{m-j}$, obtained by choosing the block and then partitioning the rest. Dividing the resulting total by $m$ and by $B_m$ gives Equation~\eqref{eq:avg}.
\end{proof}

Concretely, $\tv$ is the per-head worth an outsider should expect when every arrangement of the outsiders is treated as equally likely; we call it the \emph{benchmark}. The weights $w_j$ of Equation~\eqref{eq:avg} are so far purely combinatorial, and stationarity, by Definition~\ref{def:stat}, is a property relative to a state. The missing link is that $\tv$ has not yet been connected to the state $\pi$ that the prior induces. The next result supplies it: the weights are the occupancy in disguise, $\pi_k=k\,w_k$, and the average $\tv$ is exactly the replicator mean fitness at $\pi$.

\begin{proposition}[The occupancy $\pi$ and the benchmark]\label{prop:ref}
The occupancy of Definition~\ref{def:pi} is $\pi_k = k\,w_k = \binom{m-1}{k-1}B_{m-k}/B_m$, and the average worth of Proposition~\ref{prop:avg} is the replicator mean fitness at $\pi$:
\[
\tv \;=\; \sum_{k=1}^{m}\pi_k\, f_k \;=\; \bar f(\pi).
\]
\end{proposition}

\begin{proof}
Fix an outsider $i$. The partitions in which $i$'s block has size $k$ number $\binom{m-1}{k-1}B_{m-k}$, since one chooses $i$'s $k-1$ blockmates and then partitions the rest. Hence $\pi_k=\binom{m-1}{k-1}B_{m-k}/B_m$, which equals $k\,w_k$ by $k\binom{m}{k}=m\binom{m-1}{k-1}$. Being the distribution of the size of $i$'s block, $(\pi_k)_k$ sums to one, which is the identity $\sum_k k\binom{m}{k}B_{m-k}=mB_m$ proved in Appendix~A. Finally $\sum_k \pi_k f_k=\sum_k k w_k\, v(k)/k=\tv$.
\end{proof}

The proposition thus delivers the two faces of $\pi$ at once. As the distribution of a random outsider's block size, it is what Definition~\ref{def:pi} promised. As a point of $\Delta^{m-1}$, the expected occupancy of a uniformly drawn partition, it is a population state, the one at which stationarity will now be evaluated. By Definition~\ref{def:stat} and Proposition~\ref{prop:ref}, size $k$ is stationary at the maximum-entropy occupancy precisely when $v(k)/k=\tv$, a linear restriction on the worth vector. Collecting these restrictions, one for each size, yields the geometry on which the rule operates.

\begin{theorem}[Stationarity hyperplanes]\label{thm:main}
For $m\ge2$, size $k$ is stationary at $\pi$ if and only if the worth vector lies on the hyperplane
\begin{equation}\label{eq:thm}
\begin{gathered}
H_k=\Bigl\{v\in\R^m:\ v(k)\Bigl(\frac1k-w_k\Bigr)-\sum_{\substack{j=1\\ j\neq k}}^{m} v(j)\,w_j=0\Bigr\},\\
k=1,\dots,m,
\end{gathered}
\end{equation}
a hyperplane through the origin of worth space. Writing its normal as $n^{(k)}$, with $n^{(k)}_k=\tfrac1k-w_k$ and $n^{(k)}_j=-w_j$ for $j\neq k$, one has $n^{(k)}\!\cdot v = v(k)/k-\tv =: r_k(v)$, the \emph{residual} of size $k$.
\end{theorem}

\begin{proof}
Substitute $\tv=\sum_j w_j v(j)$ into $v(k)/k=\tv$ and separate the $j=k$ term.
\end{proof}

In other words, a size is stationary when its members earn exactly the agnostic average, and its residual measures the deviation from that average. Now, the rule to come will select the size the prior renders least remarkable. But one obstacle must be cleared first.

\begin{remark}[Degeneracy and ties]\label{rem:degen}
The family of hyperplanes $H_1,\dots,H_m$ is never in general position. Write the normals as rows of $M=D-\mathbf 1 w^{\!\top}$ with $D=\operatorname{diag}(1,\tfrac12,\dots,\tfrac1m)$, and compute the kernel of $M$ directly. $Mv=0$ says $v(k)/k=\tv$ for every $k$: writing $c$ for the common value, $v(j)=cj$ for all $j$. Conversely, if $v(j)=cj$ then $\tv=c\sum_j j\,w_j=c$ by Proposition~\ref{prop:ref}, and every residual vanishes. The kernel of $M$ is therefore exactly the \emph{additive line} $\{v(j)=cj:\ c\in\R\}$, so $\operatorname{rank}M=m-1$ and all $m$ hyperplanes intersect precisely in this line, on which every size earns the benchmark ($f_j\equiv c$) and every residual vanishes. Additive worths are thus the natural degenerate locus of the rule. Each $H_k$ is a hyperplane (its normal has the nonzero entries $-w_j$), hence Lebesgue null. For $m\ge3$ no two normals are proportional by any scalar, including $-1$: since every $w_j>0$, proportionality in the coordinates outside $\{k,l\}$ forces the scale to equal one, which the $k$th coordinate contradicts. Each locus where two hyperplane distances coincide is therefore, the residuals being linear in $v$, a union of two hyperplanes, again null. A generic worth vector therefore lies on no $H_k$ and admits a unique nearest hyperplane; at ties, including the additive line, we break towards the smallest size, and this convention is part of the rule. Two further facts are worth recording. The left kernel of $M$ is spanned by $\pi$: $\pi^{\top}M=0$, so $\sum_k\pi_k\,r_k(v)=0$ for every worth vector, the residuals either all vanish or take both signs, and the side payments of Remark~\ref{rem:interp} below balance at $\pi$. And the small cases are degenerate: for $m=2$ the two normals are opposite, $n^{(2)}=-n^{(1)}$, so the two distances coincide for every worth vector and the rule returns size one throughout, while for $m=1$ the single normal vanishes and Equation~\eqref{eq:classifier} is undefined. The rule is therefore informative only for $m\ge3$, which we assume whenever it is applied.
\end{remark}

Remark~\ref{rem:degen} settles well-definedness, and the rule can be stated. Despite its appearance, the formula below says three simple things. Compute each size's residual, scale it by the fixed length of its hyperplane's normal, and pick the size with the smallest scaled value.

\begin{definition}[The prediction rule]\label{def:rule}
Given a worth vector $v$, predict
\begin{equation}\label{eq:classifier}
\begin{split}
\hat k(v)\;&=\;\arg\min_{1\le k\le m}\
\frac{\left|\,v(k)\!\left(\tfrac1k-w_k\right)-\sum_{j\neq k}v(j)\,w_j\,\right|}
{\sqrt{\left(\tfrac1k-w_k\right)^2+\sum_{j\neq k} w_j^{\,2}}}\\[2pt]
&=\;\arg\min_{1\le k\le m}\ \frac{|r_k(v)|}{\lVert n^{(k)}\rVert},
\end{split}
\end{equation}
the size whose stationarity hyperplane lies nearest, with ties broken as in Remark~\ref{rem:degen}.
\end{definition}

\begin{remark}[What the rule minimises]\label{rem:interp}
Both quantities in Equation~\eqref{eq:classifier} have exact operational meanings, which is why the pair is of interest beyond its closed form. The raw residual $|r_k|=|v(k)/k-\tv|$ is the per-capita side payment that brings a member of a size-$k$ block exactly to the benchmark. Concretely, a transfer of $|r_k|$ per member closes the gap between what size $k$ earns and what the agnostic average pays, so the residual ranking $\arg\min_k|r_k|$, the rule's companion in Section~\ref{sec:bench}, selects the size cheapest to stabilise by side payments. The normalised distance $|r_k|/\lVert n^{(k)}\rVert$ is the smallest Euclidean perturbation of the worth vector that would make size $k$ stationary at $\pi$, being the distance from $v$ to the hyperplane $H_k$, so the rule selects the size whose stationarity is nearest to holding in the data. Both selections target stability at the agnostic state, not profit, and by Corollary~\ref{cor:classifier} below both depend on the worths only through the axis of their displacement from the additive line; Section~\ref{sec:bench} prices these choices in welfare terms.
\end{remark}

Two practical questions remain. How much does a prediction cost to compute, and does it matter in which units the worths are measured? The corollary answers both.

\begin{corollary}[Complexity and invariance]\label{cor:classifier}
Prediction is a single nearest-hyperplane computation on the observable worth vector. The input is the pair $(m,v)$, an integer and a vector of $m$ real worths, and costs are counted in arithmetic operations on these numbers. The weights $w_1,\dots,w_m$ depend only on $m$ and are computed once from $B_0,\dots,B_m$, integers of $O(m\log m)$ digits, in $O(m^2)$ arithmetic preprocessing, together with $\sum_j w_j^{\,2}$. Given $v$, the benchmark $\tv=\sum_j w_j\,v(j)$ is one $O(m)$ inner product; each residual $r_k=v(k)/k-\tv$ then costs $O(1)$, each norm in Equation~\eqref{eq:classifier} costs $O(1)$ using the precomputed $\sum_j w_j^{\,2}$, and the minimum is maintained in one pass, so a prediction costs $O(m)$ per query. The numerator of Equation~\eqref{eq:classifier} is the absolute value of a linear functional of $v$ that vanishes on the vector $\iota=(1,\dots,m)$, since $\iota$ spans the kernel of $M$ (Remark~\ref{rem:degen}), and the denominator does not depend on $v$; hence $\hat k(\lambda v+c\,\iota)=\hat k(v)$ for every $\lambda\neq0$ and $c\in\R$. The rule depends on the worths only through the axis of their displacement from the additive line; neither the size nor the sign of the displacement matters. Section~\ref{sec:bench} returns to this limitation.
\end{corollary}

By Remark~\ref{rem:degen} the smallest nondegenerate case is $m=3$, where the geometry can be drawn.

\begin{figure}[ht]
\centering
\includegraphics[width=\linewidth]{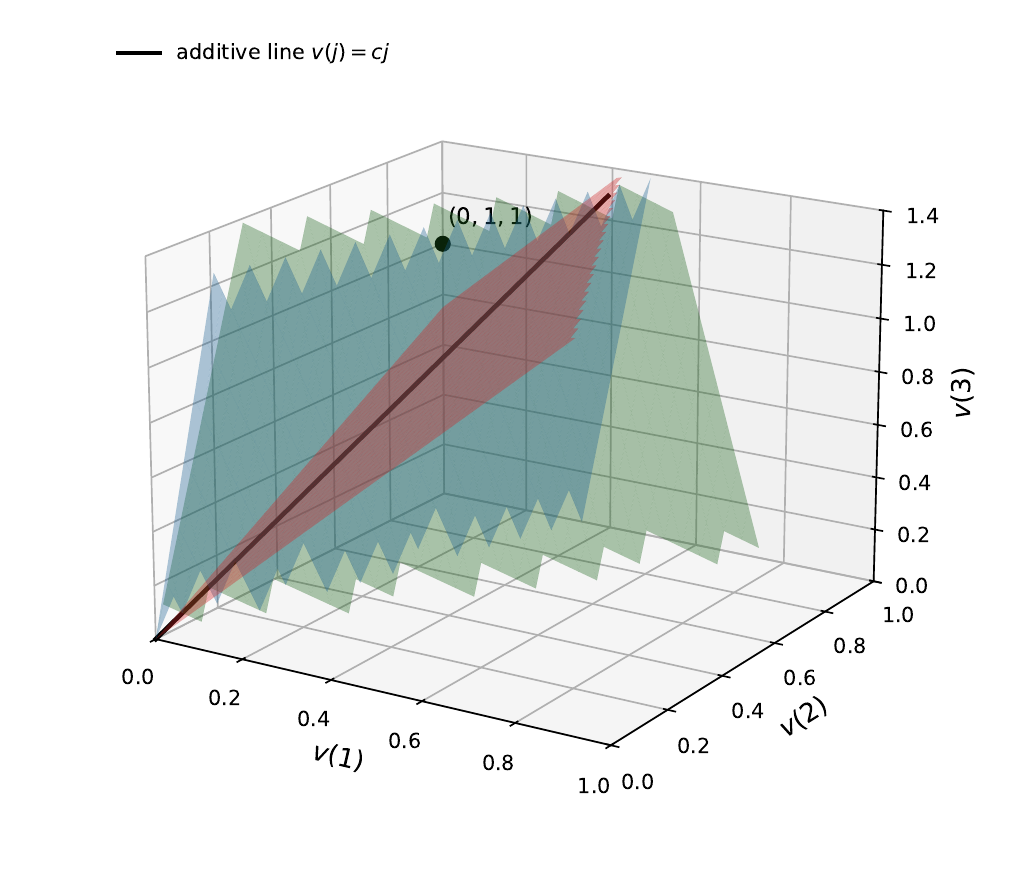}
\caption{The three stationarity hyperplanes $H_1,H_2,H_3$ for $m=3$, their common additive line $v(j)=cj$ (Remark~\ref{rem:degen}), and the worth vector $(0,1,1)$ of Example~\ref{ex:m3}. The rule predicts the size of the nearest hyperplane, here the grand coalition.}
\label{fig:planes}
\end{figure}

\begin{example}[Three outsiders]\label{ex:m3}
Let $m=3$, so $B_0=B_1=1$, $B_2=2$, $B_3=5$, and Equation~\eqref{eq:avg} gives $(w_1,w_2,w_3)=(\tfrac25,\tfrac15,\tfrac1{15})$. Multiplying Equation~\eqref{eq:thm} by $15$, the hyperplanes for sizes one, two and three are
\[
\begin{aligned}
9\,v(1)-3\,v(2)-v(3)&=0,\\
-6\,v(1)+\tfrac{9}{2}\,v(2)-v(3)&=0,\\
-6\,v(1)-3\,v(2)+4\,v(3)&=0 .
\end{aligned}
\]
For the worths $v(1)=0$, $v(2)=v(3)=1$ of [12] the distances of Equation~\eqref{eq:classifier} to $H_1,H_2,H_3$ are $0.419$, $0.463$ and $0.128$, so the nearest hyperplane is that of the grand coalition and the deviators predict the three outsiders merge (Figure~\ref{fig:planes}). The benchmark is $\tv=4/15$ with residuals $r=(-0.267,\,0.233,\,0.067)$ for sizes one, two and three, so the residual ranking $\arg\min_k|r_k|$ also selects the merger, whereas the best response $\arg\max_k v(k)/k$ selects the pair, since $v(2)/2=\tfrac12$ exceeds $v(3)/3=\tfrac13$. The construction predicts proximity to the agnostic average, not individual optimality.
\end{example}

The example read the rule geometrically, as a distance in worth space. However, the same computation also has a dynamical face.

\paragraph{Dynamical reading}
Start the replicator flow of Equation~\eqref{eq:replicator} exactly at the agnostic state $\pi$. By Proposition~\ref{prop:ref}, in the notation of Theorem~\ref{thm:main}, the initial per-capita growth rate of each size's share is its residual,
\begin{equation}\label{eq:resid}
\left.\frac{\dot x_k}{x_k}\right|_{x=\pi}=\frac{v(k)}{k}-\tv= r_k ,
\end{equation}
so the residual companion of Section~\ref{sec:bench} selects the size whose share the flow disturbs least, and the rule, its norm-adjusted variant, the size closest to stationarity in worth space when the flow is started from ignorance. This is a diagnostic of the worth function at $\pi$, not a prediction of where the flow ends, because with constant fitness the flow from interior states converges to the best response $\arg\max_k v(k)/k$ (shares evolve over time $t$ as $x_k(0)\,e^{f_k t}$, normalised, so the largest fitness dominates). Section~\ref{sec:bench} measures how often the rule's pick coincides with this limit.

Running that comparison, and every benchmark of the next section, scores a whole structure rather than a size. The size $\hat k$ is completed into one by the tiling convention of Definition~\ref{def:tiling}.

\section{Benchmarks against the welfare optimum}\label{sec:bench}

\begin{figure}[ht]
\centering
\includegraphics[width=\linewidth]{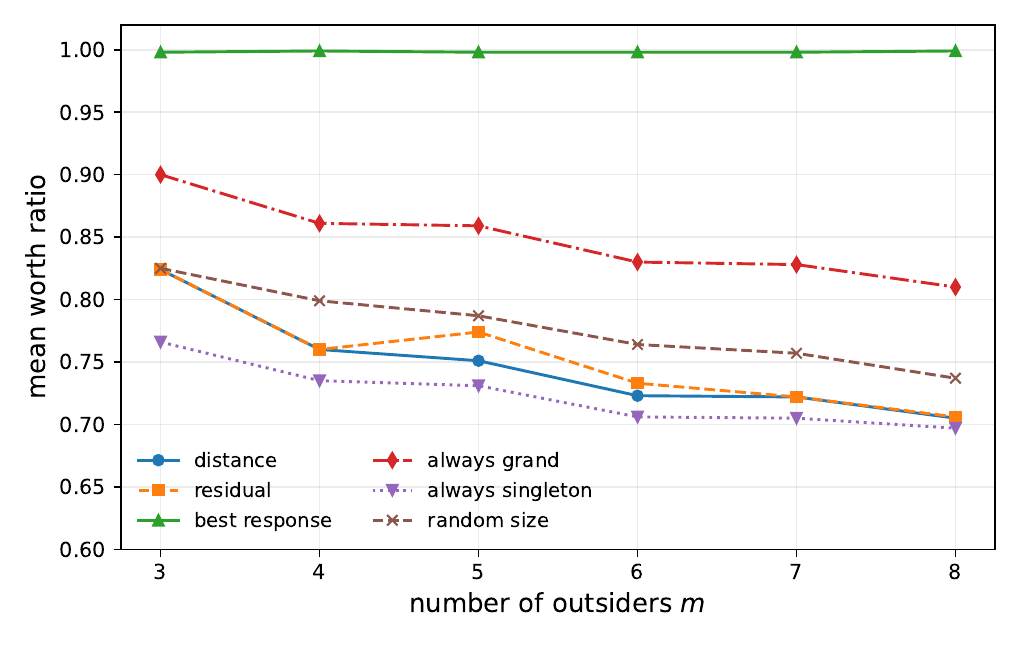}
\caption{Mean worth ratio of each selection as the number of outsiders grows (the entries of Table~\ref{tab:val}), including the random-size baseline.}
\label{fig:validation}
\end{figure}

A prediction rule needs a yardstick. Ours is not designed to maximise welfare, so the honest questions are how much welfare it forgoes, how it compares with simple fixed selections, which Remark~\ref{rem:cores} will connect to the classical conjectures of the core literature, and how often it departs from the best response. The five families below are chosen to span the qualitative shapes of returns to scale, increasing, decreasing, linear, tipping and irregular, on which the answers could differ. We benchmark six selections on five families of worth functions, for $m=3,\dots,8$. The selections are the distance rule of Equation~\eqref{eq:classifier}; the residual ranking $\arg\min_k|r_k|$ (the unnormalised companion of Definition~\ref{def:rule}, included as a robustness check on the normalisation); the best response $\arg\max_k v(k)/k$; always grand ($k=m$); always singleton ($k=1$); and a uniformly random size. The families are superadditive ($v(j)=j^{1.6}$), subadditive ($v(j)=j^{0.6}$), linear ($v(j)=j$), a tipping family with a sharp scale economy above a random threshold, and random nondecreasing worths. Each selected size is scored by the worth ratio, namely the per-capita worth of its induced structure (Definition~\ref{def:tiling}) divided by the social optimum. The optimum is the largest per-capita worth over all partitions of the $m$ outsiders, computed by exhausting the integer partitions of $m$ or, equivalently, by an $O(m^2)$ dynamic program. Under Assumptions~\ref{a:anon} to~\ref{a:equal}, welfare optimisation is easy, and the open question is coordination. Table~\ref{tab:val} and Figure~\ref{fig:validation} report means over several hundred draws per $(m,\text{family})$ cell. For scale, the selections differ in cost as well as in score: always grand, always singleton and random size are $O(1)$; best response is an $O(m)$ scan; the residual ranking and the distance rule are $O(m)$ per query after the $O(m^2)$ weight preprocessing of Corollary~\ref{cor:classifier}; and the optimum itself is an $O(m^2)$ dynamic program.

\begin{remark}[The baselines and the classical conjectures]\label{rem:cores}
The baseline selections are not arbitrary. Two embody classical conjectures about outsider behaviour from the core literature. Always singleton is the conjecture underlying the $\gamma$-core, in which the outsiders disintegrate after a deviation [13]. Always grand is the conjecture underlying the $\delta$-core, in which they merge into a single bloc [14]. The Greek letters in these names are historical labels, unrelated to the differentiation parameter $\gamma$ of Section~\ref{sec:cournot}. Best response approximates the conjecture underlying the e-core, in which the outsiders choose an efficient partition among themselves [15]; the approximation gap is exactly the remainder block, and it will be visible in Table~\ref{tab:val} as the reason the best-response row sits at $0.998$ rather than $1.000$. The worst case over feasible structures, which Definition~\ref{def:belief} adopts in Section~\ref{sec:cournot}, is in the spirit of the $\alpha$- and $\beta$-approaches [16], as [9] itself notes; and random size conjectures nothing, serving as the fully agnostic control. Table~\ref{tab:val} can therefore be read as the rule against the literature's standard conjectures about the outsiders.
\end{remark}

\begin{table}[ht]
\centering
\begin{tabular}{lccccccc}
\toprule
selection & $m{=}3$ & $m{=}4$ & $m{=}5$ & $m{=}6$ & $m{=}7$ & $m{=}8$ & overall\\
\midrule
distance        & 0.824 & 0.760 & 0.751 & 0.723 & 0.722 & 0.705 & 0.748\\
residual        & 0.824 & 0.760 & 0.774 & 0.733 & 0.722 & 0.706 & 0.753\\
best response   & 0.998 & 0.999 & 0.998 & 0.998 & 0.998 & 0.999 & 0.998\\
\midrule
always grand    & 0.900 & 0.861 & 0.859 & 0.830 & 0.828 & 0.810 & 0.848\\
always singleton& 0.766 & 0.735 & 0.731 & 0.706 & 0.705 & 0.697 & 0.723\\
random size     & 0.825 & 0.799 & 0.787 & 0.764 & 0.757 & 0.737 & 0.778\\
\bottomrule
\end{tabular}
\caption{Mean worth ratio, the per-capita worth of the induced structure divided by the social
optimum, by selection and number of outsiders $m$.}
\label{tab:val}
\end{table}

The table tells one consistent story across $m$. Best response is essentially efficient (overall ratio $0.998$; the only loss is the remainder block), since any partition's per-capita worth is a convex combination of the $f_k$ and hence at most $\max_k f_k$, and on these families the remainder loss is negligible. The rule and its residual companion pool near three quarters, $0.748$ for distance and $0.753$ for residual, a near-tie showing that the normalisation, the one discretionary step in Definition~\ref{def:rule}, barely moves the welfare outcome. We state the unfavourable comparisons plainly. Both fall below the always-grand reference (overall $0.848$) at every $m$, and the random-size baseline (overall $0.778$) exceeds both at every $m$, narrowly at $m=3$. Both remain above always-singleton (overall $0.723$) throughout, and the pair agrees with the best response, the limit of the replicator flow of Equation~\eqref{eq:replicator} from interior states (Section~\ref{sec:rule}), on a fifth of instances or fewer. The reading is that these are benchmarks, not validations. The rule and its companion are not welfare heuristics; they are diagnostics of how far a worth function displaces an uncoordinated population from its agnostic state $\pi$, and the gap to the welfare-designed baselines quantifies exactly that distinction. Part of the gap has a structural source: by the invariance of Corollary~\ref{cor:classifier} the pair cannot distinguish a worth function from its mirror image $-v+c\,\iota$, so it forgoes exactly the directional information that welfare comparisons use. The goal of the prediction, per Remark~\ref{rem:interp}, is stability at the agnostic state rather than welfare, and on that criterion each is optimal by construction: the companion minimises the stabilising side payment, the rule the stationarising perturbation of the worths. Always grand, for instance, commits to size $m$ whatever its distance from the average. Table~\ref{tab:val} prices this stance in welfare terms; it does not rank the selections on their own objectives. The two jobs are different, and the table's advice follows. A manager splitting $m$ couriers into delivery teams to maximise output per courier should use best response, the near-optimal design tool. An analyst instead holds estimated coalition worths and asks which size the market's payoff structure leaves nearest to stationarity at the agnostic state, and how near. That analyst should use the rule or its residual companion. What they return is exactly that answer, and the welfare score is beside the point. The linear family ($v(j)=j$) deserves disclosure. Its worths lie precisely on the additive line of Remark~\ref{rem:degen}, where every partition attains the optimum, so its cells equal one for all six selections and carry no discriminating information.

\section{Prediction and the Cournot $j$-belief core}\label{sec:cournot}

The benchmarks of Section~\ref{sec:bench} score welfare; they do not test the rule at its original job, standing in for the deviators' missing prediction. The coalitional-belief framework of [9] was built for exactly that situation. There, a coalition deviating in a Cournot oligopoly cannot compute the outsiders' reaction either; instead, it holds a belief about how many coalitions the outsiders will form. The framework then characterises which beliefs support cooperation. Our rule produces precisely such a belief, the number of coalitions in the structure it predicts. The two frameworks can therefore be confronted on a single question. Is the deviation deterred by a stable division of the grand coalition's worth when the outsiders react as the rule predicts? Cournot matters for a second reason as well, anticipated in Remark~\ref{rem:anon}. Its worths violate Assumption~\ref{a:ownsize}, so the section doubles as a demonstration of the reduction that carries the rule outside its home assumptions.

In [9], firms $i\in N$ produce differentiated goods with inverse demand $a-q_i-\gamma\sum_{l\neq i}q_l$ and cost $c\,q_i$, $0<c<a$, with differentiation $\gamma\in(-1,1]$: substitutes for $\gamma>0$, complements for $\gamma<0$, homogeneous goods at $\gamma=1$. For complements we impose condition $K:\ \gamma>-1/(n-1)$, which guarantees existence of the Cournot equilibrium, and the grand coalition's monopoly worth is $v(N)=n(a-c)^2/\bigl(4(1+\gamma(n-1))\bigr)$ [9]. \begin{definition}[Cournot partition-function game]\label{def:cournotpfg}
Given a coalition structure $Q$ of $N$, each block coordinates its members' quantities and the blocks play a Cournot--Nash equilibrium among themselves; $u(b;Q)$ is block $b$'s total equilibrium profit. The pair $(N,u)$ is the \emph{Cournot partition-function game}. It is anonymous, so Assumption~\ref{a:anon} holds with $u(b;Q)=v(|b|;\lambda(Q))$, but a block's worth varies with the entire size profile, so Assumption~\ref{a:ownsize} fails.
\end{definition}

For reproducibility we record the equilibrium block profits in closed form, the block-symmetric form of the equilibrium computed in [9] for the deviating coalition. Writing $A=a-c$, $\alpha_t=1+\gamma(t-1)$ and $\theta_t=t/\bigl(2(1-\gamma)+\gamma t\bigr)$, a block of size $t$ in a structure with block sizes $t_1,\dots,t_r$ earns
\[
u \;=\; A^{2}\,\frac{\alpha_t}{t}\,\frac{\theta_t^{2}}{(1+\gamma\Theta)^{2}},\qquad \Theta=\sum_{b=1}^{r}\theta_{t_b}.
\]
The formula gives the interior equilibrium for $\gamma\in(-1,1]$ under condition $K$; at $\gamma=1$ the block profits remain determined although the split within a block is not. All grid computations below use it, with the Bell-number weights of Equation~\eqref{eq:avg} computed in exact integer arithmetic.

The framework's primitive is the belief itself, formalised as follows.

\begin{definition}[$j$-belief worth and core {[9]}]\label{def:belief}
A \emph{$j$-belief} assigns to the deviating coalition $S$ an integer $j(s)\in\{1,\dots,n-s\}$: $S$ believes the outsiders will reorganise into \emph{at least} $j(s)$ coalitions. The $S$-feasible partitions are $P(j,S)=\{P \text{ of } N\setminus S:\ |P|\ge j(s)\}$, the $j$-belief worth is
\[
v^{j}(S)\;=\;\min_{P\in P(j,S)} v^{P}(S),\qquad v^{j}(N)=v(N),
\]
where $v^{P}(S)=u(S;Q_P)$ is $S$'s equilibrium profit against the outsider structure $P$ (Definition~\ref{def:cournotpfg}), and the \emph{$j$-belief core} is the core of $(N,v^{j})$.
\end{definition}

Read plainly, a $j$-belief says that however the outsiders react they will split into at least $j(s)$ groups, and the deviating coalition prepares for the worst case consistent with that. This worst case is the $\alpha$- and $\beta$-spirit of Remark~\ref{rem:cores}.

For substitutes, coalition formation among the outsiders favours the deviators, since coarser structures raise $S$'s equilibrium profit [9]. The minimum in Definition~\ref{def:belief} is therefore attained at maximal fragmentation and, the feasible sets being nested, is the same for every $j$; the belief bites not through the worth but through the deterrence conditions, which must hold uniformly across $P(j,S)$ and therefore bind at the coarsest feasible scenario, exactly $j(s)$ coalitions. The threshold results quoted next are of this uniform kind: facts (i) and (ii) of Lemma~\ref{lem:facts} locate the worth extremes over compositions, facts (iii) and (iv) give the deterrence thresholds, and the propositions that follow compare the rule's prediction with those thresholds.

\begin{lemma}[{Facts from [9]}]\label{lem:facts}
\emph{(i)} ([9], Prop.~1) For $\gamma\in(0,1)$ and fixed $j$, over structures with $j$ outsider coalitions $S$'s worth is minimised at the approximately even split and maximised when all coalitions but one are singletons.
\emph{(ii)} ([9], Cor.~2) For $\gamma=1$ the worth against $j$ outsider coalitions depends only on their number: $v^{P_j}(S)=\bigl((a-c)/(j+2)\bigr)^2$.
\emph{(iii)} ([9], Cor.~2) For $\gamma=1$ the equal split of $v(N)$ deters a size-$s$ deviation against $j$ outsider coalitions if and only if $j\ge\lceil 2(\sqrt{n/s}-1)\rceil$; the game being symmetric, the equal split is the pivotal allocation for core nonemptiness.
\emph{(iv)} ([9], Thms.~1 and~2) For complements, that is $\gamma\in(-1,0)$ under condition $K$, every $j$-belief core is nonempty; for substitutes with $\gamma\in(0,1)$ the $j$-belief core is nonempty provided $j(s)>j^{*}(n,s,\gamma)$ for each deviating size $s$, where
\[
j^{*}(n,s,\gamma)\;=\;\frac{2}{\gamma}\Bigl[\sqrt{\bigl(1+\gamma(n-1)\bigr)\bigl(1+\gamma(s-1)\bigr)}\;-\;\bigl(1+\gamma(s-1)\bigr)\Bigr].
\]
(The source writes $j^{*}(n,\gamma)$; its formula depends on $s$, which our notation makes explicit.)
\end{lemma}

The prediction rule connects to this framework through the number of coalitions it induces, and the connection is a one-line derivation rather than an imported result. Applying Equation~\eqref{eq:classifier} to the outsiders' worth vector yields $\hat k$; the tiling convention of Definition~\ref{def:tiling} completes $\hat k$ to a structure with $\lceil(n-s)/\hat k\rceil$ outsider coalitions; and since a $j$-belief is precisely a number of outsider coalitions, this count is itself an admissible belief, lying in $\{1,\dots,n-s\}$ because $1\le\hat k\le n-s$:
\[
\beta(v)\;=\;\bigl\lceil (n-s)/\hat k\bigr\rceil .
\]
This belief corresponds to a structure whose blocks all have size $\hat k$ except at most one remainder, and it carries the count rather than the composition. For $\gamma=1$ any composition with the same count is worth-equivalent (Lemma~\ref{lem:facts}(ii)), so the structural consistency between the two frameworks is exact there. Whether the predicted deviation is deterred then reduces to comparing $\beta(v)$ with the thresholds of Lemma~\ref{lem:facts}.

\begin{proposition}[Exact deterrence at $\gamma=1$]\label{prop:homcons}
For homogeneous goods ($\gamma=1$), the equal split of $v(N)$ deters the size-$s$ deviation against every outsider structure with at least $\beta(v)$ coalitions if and only if
$\bigl\lceil(n-s)/\hat k\bigr\rceil\ge\bigl\lceil 2(\sqrt{n/s}-1)\bigr\rceil$.
\end{proposition}

\begin{proof}
By Lemma~\ref{lem:facts}(ii) the deviators' worth falls as the outsiders fragment, so the binding feasible scenario has exactly $\beta(v)$ coalitions; apply Lemma~\ref{lem:facts}(iii) with $j=\beta(v)$.
\end{proof}

For homogeneous goods the criterion above is exact. Away from that boundary no exact criterion is available, but we can still ask how fast the sufficiency threshold $j^{*}$ grows with the number of firms, and the answer is clean.

\begin{proposition}[Growth of the sufficiency threshold]\label{prop:diffcons}
Fix $s$ and $\gamma\in(0,1)$ and write $\sigma=1+\gamma(s-1)$. Then
$j^{*}(n,s,\gamma)=\dfrac{2\sqrt{\sigma}}{\sqrt{\gamma}}\,\sqrt n+O(1)$ as $n\to\infty$, and for large $n$ the deterrence condition $\beta(v)>j^{*}(n,s,\gamma)$ of Lemma~\ref{lem:facts}(iv) holds exactly when
$\hat k<\bigl(1+o(1)\bigr)\dfrac{\sqrt\gamma}{2\sqrt\sigma}\,\sqrt n$; equivalently, deterrence via Lemma~\ref{lem:facts}(iv) requires the outsiders to fragment into at least order-$\sqrt n$ coalitions.
\end{proposition}

\begin{proof}
With $\nu=1+\gamma(n-1)$, $j^{*}=\frac2\gamma(\sqrt{\nu\sigma}-\sigma)$. As $n\to\infty$, $\sqrt{\nu}=\sqrt{\gamma}\sqrt n+O(n^{-1/2})$, so $j^{*}=\frac{2\sqrt{\sigma}}{\sqrt{\gamma}}\sqrt n+O(1)$; the size criterion restates $\beta(v)>j^{*}$ with $\beta(v)=\lceil(n-s)/\hat k\rceil$.
\end{proof}

One caution is in order before using this threshold numerically. The square-root rate in Proposition~\ref{prop:diffcons} is reliable, but the constant in front of it is not, and the next remark explains why.

\begin{remark}[The bound is not tight at $\gamma=1$]\label{rem:tight}
Lemma~\ref{lem:facts}(iv) is a sufficiency statement, and [9] itself notes the slack at the homogeneous boundary. Evaluating the threshold at $\gamma=1$ gives $j^{*}(n,s,1)=2(\sqrt{ns}-s)\ge 2(\sqrt{n/s}-1)$, the exact criterion of Lemma~\ref{lem:facts}(iii). The two differ by a factor of $s$ (for instance $32$ against $8$ at $(n,s)=(100,4)$). The $\Theta(\sqrt n)$ growth is what Proposition~\ref{prop:diffcons} uses; the constant should not be read as sharp.
\end{remark}

The threshold side of the comparison is now in place. What remains is the rule's side. Propositions~\ref{prop:homcons} and~\ref{prop:diffcons} took the worth vector fed to the rule as given, and for Cournot worths that vector must first be constructed.

\begin{figure}[ht]
\centering
\includegraphics[width=\linewidth]{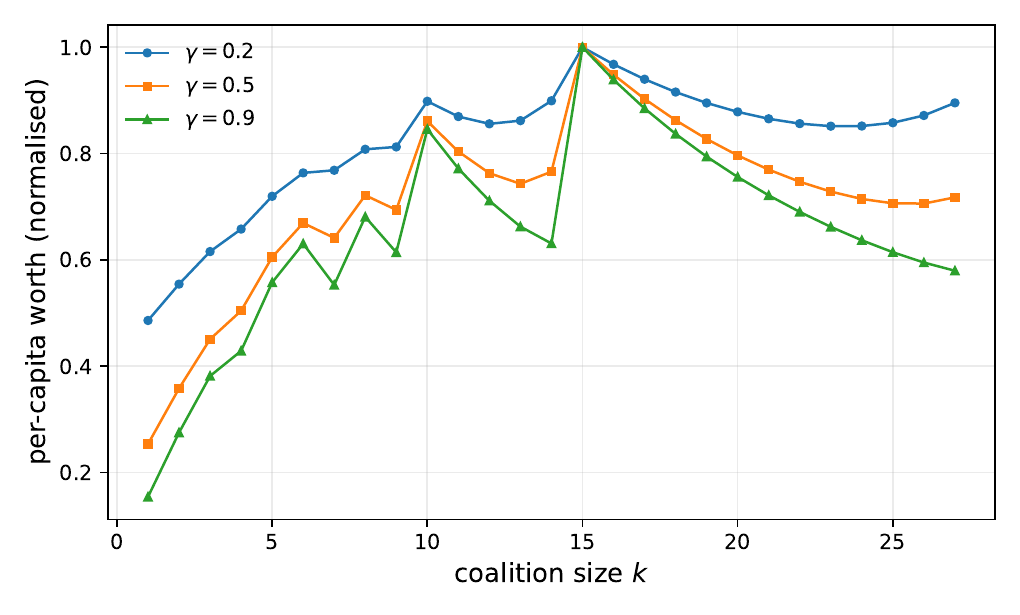}
\caption{Self-consistent per-capita worth $v(k)/k$ of an outsider coalition (normalised) against its own size, for $n=30$, $s=3$ and $\gamma\in\{0.2,0.5,0.9\}$. Under this construction the prediction tracks the prior's typical block size rather than the peak, and in particular stays far from the fine fragmentation that the sufficiency threshold for substitutes asks for. At the figure's own parameters the exact deterrence test nonetheless passes for all three $\gamma$.}
\label{fig:percapita}
\end{figure}

\paragraph{Self-consistent worths and the induced belief}
To apply the rule when Assumption~\ref{a:ownsize} fails, we reduce the Cournot worth to own-size form by self-consistency. An outsider contemplating a size-$k$ coalition evaluates the equilibrium worth $v(k)$ of that coalition when the whole population, the deviators included, tiles into size-$k$ blocks, the remainder pooled into one block. Informally, each outsider asks the same question for every $k$: if everyone organised into teams of my size, what would my team earn? The retiling includes the deviators so that a single worth curve serves every deviating size $s$; holding $S$ fixed as a block instead is equally admissible, and it changes predictions materially at strong substitution. On the same grid it yields $\hat k$ up to $27$, and the exact test below then passes in $453$ of the $564$ cases, against $488$ under our construction, with failures no longer confined to weak substitution. We flag the choice as a modelling one and scope what follows to it. The hypothetical serves only to price the sizes; in the deterrence analysis below, $S$ remains a fixed block and, as in Section~\ref{sec:model}, nothing about it is predicted. The resulting per-capita worth is non-monotone with an interior peak (Figure~\ref{fig:percapita}); the prediction itself, however, tracks the prior's typical block size under this construction, $\hat k$ between two and four throughout the grid, so the induced number of coalitions grows roughly linearly in $n$. Across $\gamma\in\{0.2,0.5,0.9\}$, $s\in\{3,5\}$ and $s+3\le n\le100$ the induced $\beta(v)$ fell short of the sufficiency threshold $j^{*}(n,s,\gamma)$ in every case; over this range the threshold, of order $\sqrt n$ with a sizeable constant, still exceeds the roughly linear $\beta\approx(n-s)/\hat k$. Two qualifications keep the finding honest. First, the threshold is sufficient, not necessary, and it is loose away from the homogeneous boundary; tested exactly, against the structure most favourable to the deviators among those with at least $\beta(v)$ coalitions, the equal split deters the deviation in $488$ of the $564$ grid cases, including every case with $\gamma\in\{0.5,0.9\}$; all failures occur at $\gamma=0.2$. Second, the shortfall is a moderate-$n$ phenomenon: because $\hat k$ grows slowly, Proposition~\ref{prop:diffcons} implies that $\beta$ eventually outruns $j^{*}$ whenever $\hat k=o(\sqrt n)$, and the grid is consistent with this. For complements deterrence is automatic by Lemma~\ref{lem:facts}(iv); for homogeneous goods Proposition~\ref{prop:homcons} is the exact criterion.

\section{Concluding remarks}\label{sec:conc}

The construction leaves three questions open. First, the axiomatic status of the prior: exchangeability fixes only the within-profile symmetry, maximum entropy selects uniformity over the labelled partitions, uniformity over size profiles induces different weights in Equation~\eqref{eq:avg}, and a characterisation of either choice from primitive invariance axioms is missing. Second, behavioural foundations: identifying learning or perturbed dynamics whose quasi-stationary behaviour selects the nearest-stationarity size would convert the diagnostic of Equation~\eqref{eq:resid} into a prediction. Third, relaxing Assumption~\ref{a:ownsize} directly, replacing the self-consistency device of Section~\ref{sec:cournot} by fixed-point occupancies, would extend the rule to fully externality-laden anonymous games.

\appendix
\section{A Bell-number identity}\label{app:identity}
For $m\ge1$, $\sum_{k=1}^{m} k\binom{m}{k}B_{m-k}=m\sum_{k=1}^{m}\binom{m-1}{k-1}B_{m-k}
=m\sum_{i=0}^{m-1}\binom{m-1}{i}B_{i}=m\,B_m$, using $k\binom mk=m\binom{m-1}{k-1}$, the substitution
$i=m-k$, the symmetry $\binom{m-1}{m-1-i}=\binom{m-1}{i}$, and the Bell recurrence. This is the
statement that the reference distribution $\pi$ of Proposition~\ref{prop:ref} sums to one.

\section*{Declaration of competing interest}
The author declares that he has no known competing financial interests or personal relationships
that could have appeared to influence the work reported in this paper.

\section*{Data availability}
A Python script reproducing all figures and the grid computations of Section~\ref{sec:cournot} is
included as supplementary material. No external data are used.

\end{document}